%% file: monophoton_paper2015.tex
\pdfoutput=1 

\documentclass[cernpreprint,texlive=2011,txfonts,UKenglish]{atlasdoc}

\usepackage{atlaspackage} 
\usepackage{atlasphysics} 
\usepackage{atlasbiblatex}
\usepackage{monophoton_paper2015-defs}

\AtlasTitle{A standalone cover for CERN preprints}
\PreprintIdNumber{CERN-EP-2016-060}
\AtlasJournalRef{JHEP06(2016)059}
\AtlasDOI{10.1007/JHEP06(2016)059}
\AtlasAbstract{ %
\input{Abstract}

}

\AtlasTitle{Search for new phenomena in events with a photon and missing transverse momentum in $pp$ collisions at $\sqrt{s}=13\TeV$ with the ATLAS detector}
\author{The ATLAS Collaboration}

\hypersetup{pdftitle={ATLAS draft},pdfauthor={The ATLAS Collaboration}}

\begin{document}
\maketitle

\tableofcontents
\clearpage

\section{Introduction}
\label{sec:intro}
\input{Introduction}

\section{The ATLAS detector}
\label{sec:detector}
\input{Detector}

\section{Monte Carlo simulation samples}
\label{sec:simsample}
\input{Mcarlo-samples}

\section{Event reconstruction}
\label{sec:Objects}
\input{Objects}

\section{Event selection}
\label{sec:Selection}
\input{Event-selection}

\section{ Background estimation}
\label{sec:BGEstimation}
\input{Background.tex}

\section{Results}
\label{sec:result}
\input{Results.tex}

\section{ Systematic uncertainties} 
\label{sec:syst}
\input{Systematics.tex}

\section{Interpretation of results}
\label{sec:interpretation}
\input{Interpretation.tex}

\FloatBarrier

\section{Conclusion}
\label{sec:conclusion}
\input{Conclusion}

\section*{Acknowledgements}
\input{Acknowledgements}

\printbibliography

\clearpage


\input{atlas_authlist}

\end{document}

%% file: Abstract.tex
Results of a search for new phenomena in events with an energetic photon and large missing transverse momentum with the ATLAS experiment 
at the Large Hadron Collider are reported. The data were collected in proton--proton collisions at a centre-of-mass energy of 13 \TeV~ and correspond to an integrated  luminosity of 3.2 \ifb.  
The observed data are in agreement with the Standard Model expectations.  
Exclusion limits are presented in models of new phenomena including pair production of dark matter candidates or large extra spatial 
dimensions.
In a simplified model of dark matter and an axial-vector mediator, 
the search excludes mediator masses below $710\GeV$ for dark matter candidate masses below $150\GeV$. 
In an effective theory of dark matter production, 
values of the suppression scale $M_*$ up to $570\GeV$ are excluded and the effect of truncation for various coupling values is reported.
For the ADD large extra spatial dimension model the search places more stringent limits than earlier searches in the same event topology, 
excluding $M_{\rm D}$ up to about 2.3 (2.8)$\TeV$ for two (six) additional spatial dimensions; 
the limits are reduced by 20--40\% depending on the number of  additional spatial dimensions when applying a truncation procedure.

%% file: Introduction.tex
Theories of dark matter (DM) or large extra spatial dimensions (LED) 
predict the production of events that contain a high transverse momentum (\pt) photon and large missing transverse momentum
(referred to as $\gamma+\missET$ events) in $pp$ collisions at a higher rate than is expected in the Standard Model (SM).
A sample of $\gamma+\missET$ events with a low expected contribution from SM processes provides powerful sensitivity to models
of new phenomena~\cite{Aaltonen:2008hh,Abazov:2008kp,Abbiendi:2000hh,Achard:2003tx,Abdallah:2003np}. 

 The ATLAS~\cite{Aad:2012fw, PRD2015} and CMS~\cite{Chatrchyan:2012tea, CMS_8TeV} collaborations
 have reported limits on various models based on searches for an excess in
 $\gamma+\missET$ events using $pp$ collisions at centre-of-mass energies of $\sqrt{s}=7$~and $8 \TeV$ (LHC Run 1). This paper reports the results of a search for new phenomena in $\gamma+\missET$ events in $pp$ collisions at $\sqrt{s}=13\TeV$.

Although the existence of  DM is well established~\cite{Bertone:2004pz}, 
it is not explained by current theories.
 One candidate is a weakly interacting massive particle (WIMP, also denoted by 
$\chi$), which has an interaction strength with SM particles near the level of the weak interaction. If WIMPs interact with quarks via a 
mediator particle, they could be pair-produced in $pp$ collisions at sufficiently high energy. The $\chi\bar{\chi}$ pair would be invisible to the detector, but $\gamma+\missET$ events can be produced  via radiation of an initial-state photon in $q\bar{q}\rightarrow\chi\bar{\chi}$ interactions~\cite{Goodman:2010ku}.

\begin{figure}[htbp]
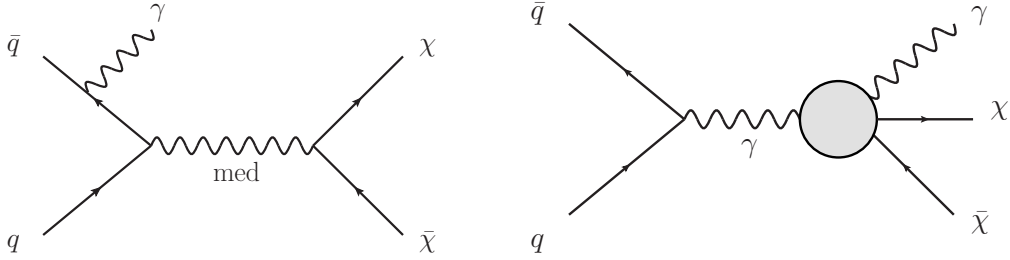
\centering
 \includegraphics[width=0.35\linewidth]{{{fig_01a}}}
 \quad \quad \quad
 \includegraphics[width=0.39\linewidth]{{{fig_01b}}}
 \caption{Production of pairs of dark matter particles ($\chi\bar{\chi}$) via an explicit $s$-channel mediator, med  (left) 
  and production of pairs of dark matter particles ($\chi\bar{\chi}$) via an effective  $\gamma\gamma\chi\bar{\chi}$ vertex (right).}
 \label{fig:Feynman_DM}
\end{figure}

A model-independent approach to dark matter production in $pp$ collision is through 
effective field theories (EFT) with various forms of interaction between the WIMPs and the SM particles \cite{Goodman:2010ku}.
However, as the typical momentum transfer in $pp$ collisions at the LHC could reach the cut-off scale 
required for the EFT approximation to be valid, 
it is crucial to  present the results of the search in terms of models that involve the explicit production of the intermediate state, as shown in Fig.~\ref{fig:Feynman_DM} (left). 
This paper focuses on simplified models assuming Dirac fermion DM candidates produced via an $s$-channel mediator with axial-vector 
interactions~\cite{Abdallah, Buchmueller, dmforum}.
In this case, the interaction is effectively described by five parameters:
the WIMP mass $m_\chi$, the mediator mass $m_{\rm med}$, the width of the mediator $\Gamma_{\rm med}$, 
 the coupling of the mediator to quarks $g_q$, and the coupling of the mediator to the dark matter particle $g_{\chi}$.
In the limit of large mediator mass, these simplified models map onto the EFT operators, with the suppression scale\footnote{The suppression scale, also referred to as $\Lambda$, is the effective mass scale of particles that are integrated out in an 
EFT. The non-renormalizable operators are suppressed by powers of 1/$\Mstar$.} 
$\Mstar$ linked to $m_{\rm med}$ by the relation $\Mstar=m_{\rm med}/\sqrt{g_qg_\chi}$~\cite{Busoni}.  

The paper also considers a specific EFT benchmark, for which neither a simplified model completion
nor the simplified models yielding similar kinematic distributions are implemented in an event generator~\cite{Crivellin}.
A dimension-7 EFT operator with direct couplings between DM and electroweak (EW) bosons, and  
describing a contact interaction of type $\gamma\gamma\chi\bar{\chi}$, is used~\cite{dmforum}.
The effective coupling to photons is parameterized by the coupling strengths $k_1$ and $k_2$, which control the strength of the coupling 
to the U(1) and SU(2) gauge sectors of the SM, respectively. In this model, dark matter production proceeds via 
$q \bar{q} \rightarrow \gamma \rightarrow \gamma\chi\bar{\chi}$, 
without requiring initial-state radiation. The process is shown in Fig.~\ref{fig:Feynman_DM} (right). 
There are four free parameters in this model: the EW coupling strengths $k_1$ and $k_2$, $m_{\chi}$, and the suppression scale $\Lambda$.

\begin{figure}[htbp]
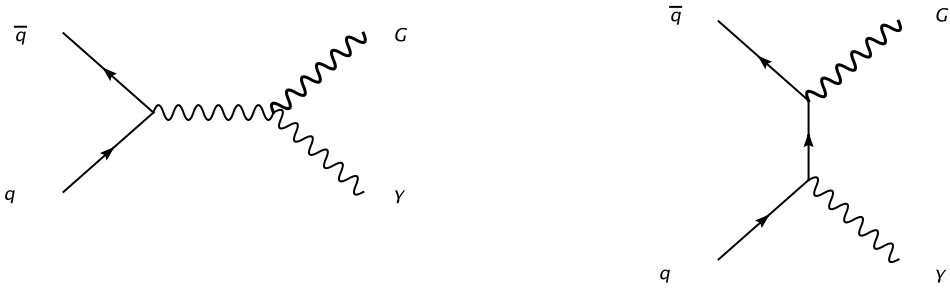
\centering
\includegraphics[width=0.49\linewidth]{{{fig_02a}}}
\includegraphics[width=0.49\linewidth]{{{fig_02b}}}
\caption{Graviton ($G$) production in models of large extra dimensions.}
\label{fig:Feynman_ADD}
\end{figure}

The ADD model of LED~\cite{ArkaniHamed:1998rs} aims to solve the hierarchy problem by 
hypothesizing the existence of $n$ additional spatial dimensions of size $R$, leading to a new fundamental scale $M_{\rm D}$ related to 
the Planck mass,  $M_{\rm Planck}$, through $M_{\rm Planck}^2 \approx M_{\rm D}^{2+{n}}R^{n}$. If these dimensions are compactified, a 
series of massive graviton ($G$) modes results.  Stable gravitons would be invisible to the ATLAS detector, but if the graviton couples to 
photons and is produced in association with a photon, the detector signature is a $\gamma+\missET$ event. Examples of graviton production are  
illustrated in  Fig.~\ref{fig:Feynman_ADD}.

The search follows a strategy similar to the search performed using the $8\TeV$ data collected during  the LHC Run 1~\cite{PRD2015} . 
Due to the increased centre-of-mass energy, the search presented 
here achieves better sensitivity for the ADD model case where direct comparison
with the $8\TeV$ search result is possible, as is shown later.
Different  DM models, proposed in Ref.~\cite{dmforum}, are also considered.

The paper is organized as follows.
A brief description of the ATLAS detector is given in Section \ref{sec:detector}. 
The signal and background Monte Carlo (MC) simulation samples used are described in Section \ref{sec:simsample}.
The reconstruction of physics objects is explained in Section \ref{sec:Objects}, and
the event selection is described in Section \ref{sec:Selection}.
Estimation of the SM backgrounds is outlined in Section \ref{sec:BGEstimation}. 
The results  are described in Section \ref{sec:result} and the systematic uncertainties are given in Section \ref{sec:syst}.
The interpretation of results in terms of  models of new phenomena including pair production of dark matter candidates or large extra spatial dimensions is described in  Section \ref{sec:interpretation}. 
A summary is given in Section \ref{sec:conclusion}.

%% file: Detector.tex
The ATLAS detector \cite{ATLAS} is a multi-purpose particle
physics apparatus with a forward-backward symmetric
cylindrical geometry and near 4$\pi$ coverage in solid angle.\footnote{ATLAS uses a right-handed coordinate system with 
its origin at the nominal  interaction point (IP) in the centre of the detector and the $z$-axis along the beam pipe. The $x$-axis points from the IP to the centre of the LHC ring, and the $y$-axis points upward. Cylindrical coordinates ($r,\phi$) are used in the transverse plane, $\phi$ being the azimuthal angle around the beam pipe. The pseudorapidity is defined in terms of the polar $\theta$ angle as ${\eta = -\ln\left[\tan(\theta/2)\right]}$.}
The inner tracking detector (ID) covers the
pseudorapidity range $|\eta| <$ 2.5, and consists of a silicon
pixel detector, a silicon microstrip detector, and,
for $|\eta| <$ 2.0, a straw-tube transition radiation tracker (TRT). 
During the LHC shutdown in 2013--14, 
 an additional inner pixel layer, known as the insertable B-layer~\cite{IBL}, 
 was added around a new, smaller radius beam pipe.
The ID is surrounded by a thin superconducting solenoid providing
a 2 T magnetic field.
 A high-granularity lead/liquid-argon sampling electromagnetic
 calorimeter covers the region $|\eta|<$~3.2 and
 is segmented longitudinally in shower depth.
The first layer, with high granularity in the $\eta$ direction, is designed to allow efficient discrimination between single photon showers and two overlapping photons originating from a $\pi^0$ decay.
The second layer collects most of the energy deposited in the calorimeter in electromagnetic showers initiated by electrons or photons.
Very high energy showers can leave significant energy deposits in the third layer, which can also be used to correct for energy leakage beyond the EM calorimeter.
A steel/scintillator-tile calorimeter provides hadronic coverage in the range $|\eta| <$ 1.7.
The liquid-argon technology is also used for the hadronic calorimeters in the end-cap region 1.5~$<|\eta|<$~3.2 and for  electromagnetic and hadronic
measurements in the forward region up to $|\eta|=$~4.9.
The muon spectrometer (MS) surrounds
the calorimeters. It consists of three large air-core superconducting toroidal magnet systems, precision tracking chambers providing 
accurate muon tracking out to $|\eta|$~=~2.7, and fast detectors for triggering in the region $|\eta| <$~2.4. 
A two-level trigger system is used to select events for offline analysis~\cite{Trigger_new}.

%% file: Mcarlo-samples.tex
Several MC simulated samples are used to estimate the signal acceptance, the detector efficiency and to help in the estimation of the SM background contributions.

For all the DM samples considered here, the values of the free parameters and the event generation settings were chosen following the recommendations given in Ref.~\cite{dmforum}.

Samples of DM production in simplified models are generated via an $s$-channel mediator with axial-vector interactions. 
The $g_q$ coupling is set to be universal in quark flavour and equal to 0.25,
$g_{\chi}$ is set to 1.0, and $\Gamma_{\rm med}$ is computed as the minimum width allowed given the couplings and masses.
A grid of points in the $m_{\chi}$--$m_{\rm med}$ plane is generated.  
The parton distribution function (PDF) set used is NNPDF30\_lo\_as\_0130~\cite{Ball:2014uwa}. 
The program \textsc{MG5}\_a\textsc{MC@NLO} v2.2.3~\cite{Alwall:2014hca} is used to generate the events, 
in conjunction with $\pythia$ 8.186 \cite{Sjostrand:2007gs} with the NNPDF2.3LO PDF set~\cite{Ball:2012cx, Ball:2013hta} and the 
A14 set of tuned parameters (tune)~\cite{ATL-PHYS-PUB-2014-021}. A photon with at least $130\GeV$ of transverse momentum 
is required in 
\textsc{MG5}\_a\textsc{MC@NLO}.
For a fixed $m_{\chi}$, higher $m_{\rm med}$ leads to harder $\pT$ and $\met$ spectra. 
For a very heavy mediator ($\geq 10\TeV$), EFT conditions are recovered.

For DM samples from an EFT model involving dimension-7 operators with a contact interaction of type $\gamma\gamma\chi\bar{\chi}$,
the parameters which only influence the cross section are set to $k_1=k_2=1.0$ and $\Lambda=3.0\TeV$. A scan over a range of values of $m_{\chi}$ is performed. The settings of the generators, PDFs, underlying-event tune and generator-level requirements are the same as for the simplified model DM sample generation described above.

Signal samples for ADD models are simulated with the $\pythiaeight$ generator, 
using the NNPDF2.3LO PDF with the A14 tune. 
A requirement of $\hat{p}_{Tmin} > 100\GeV$, where $\hat{p}_{Tmin}$ defines the lowest transverse momentum used for the generation, is applied to the leading-order (LO) matrix elements for the $2\rightarrow 2$ process to increase the efficiency of event generation.
Simulations are run for two values 
of the scale parameter $M_{\rm D}$ ($2.0$ and $3.0\TeV$) and with the number of extra dimensions, $n$, varied from two to six.

For $W/Z\gamma$ backgrounds, events containing a charged lepton and neutrino or a lepton pair 
(lepton is an e, $\mu$ or $\tau$), 
together with a photon and associated jets are simulated using the $\sherpa$ 2.1.1 generator~\cite{Gleisberg:2008ta}. 
The matrix elements including all diagrams with three electroweak couplings are calculated 
with up to three partons at LO and merged with $\sherpa$ parton shower~\cite{Schumann:2007mg} using the ME+PS@LO prescription 
\cite{Hoeche:2009rj}. The CT10 PDF set~\cite{Lai:2010vv} is used in conjunction with a dedicated parton shower tuning developed by the 
$\sherpa$ authors. 
For $\gamma^*/Z$ events with the $Z$ decaying to charged particles a requirement on the dilepton invariant mass of $m_{\ell\ell} > 10 \GeV$ 
is applied at generator level.

Events containing a photon with associated jets are also simulated using $\sherpa$ 2.1.1, generated in several bins of photon $\pt$  
from $35\GeV$ up to larger than $4\TeV$. The matrix elements are calculated at LO with up to three partons (lowest $\pt$ slice) or four partons and 
merged with $\sherpa$ parton shower using the ME+PS@LO prescription. The CT10 PDF set is used in conjunction with the  
dedicated parton shower tuning.

For $W/Z$+jets backgrounds, events containing $W$ or $Z$ bosons with associated jets are again simulated using $\sherpa$ 2.1.1. The matrix elements are calculated for up to two partons at NLO and four partons at LO using the Comix~\cite{Gleisberg:2008fv} and OpenLoops~\cite{Cascioli:2011va} matrix element generators and merged with $\sherpa$ parton shower using the ME+PS@NLO prescription~\cite{Hoeche:2012yf}. 
As in the case of the $\gamma$+jets samples, the CT10 PDF set is used together with the dedicated parton shower tuning. The $W/Z$+jets events are normalized to NNLO cross sections~\cite{ATL-PHYS-PUB-2016-003}. These samples are also generated in several $\pt$ bins.

Multi-jet processes are simulated using the $\pythia$ 8.186 generator. The A14 tune is used together with the NNPDF2.3LO PDF 
set. The EvtGen v1.2.0 program~\cite{EvtGen} is used 
to simulate the bottom and charm hadron decays. 

Diboson processes with four charged leptons, three charged leptons and one neutrino or two charged leptons and two neutrinos are simulated using the $\sherpa$ 2.1.1 generator. The matrix elements contain all diagrams with four electroweak vertices. They are calculated for up to one parton 
(for either four charged leptons or two charged leptons and two neutrinos) or zero partons (for three charged leptons and one neutrino) at NLO, and up to three partons at LO using the 
Comix and OpenLoops matrix element generators and merged with $\sherpa$ parton shower using the ME+PS@NLO prescription. The CT10 PDF set is used in conjunction with the dedicated parton shower tuning. The generator cross sections are used in this case, which are at NLO. 

For the generation of $t\bar{t}$ and single top quarks in the $Wt$ and $s$-channel, the \textsc{Powheg-Box} v2~\cite{Alioli:2010xd, Alioli:2009je}
generator is used, with the CT10 PDF set used in the matrix element calculations. For all top processes, top-quark spin correlations 
are preserved. For $t$-channel production, top quarks are decayed using MadSpin~\cite{Artoisenet:2012st}. The parton shower, fragmentation, and 
the underlying event are simulated using $\pythia$~6.428~\cite{pythia} with the CTEQ6L1~\cite{Pumplin:2002vw} PDF sets and the corresponding 
Perugia 2012 tune~\cite{Skands:2010ak}. The top mass is set to $172.5\GeV$. The EvtGen v1.2.0 program is used for properties of the bottom 
and charm hadron decays. 

Multiple $pp$  interactions in the same or neighbouring bunch crossings
superimposed on the hard physics process (referred to as  pile-up) 
are simulated with the soft QCD processes of $\pythia$ 8.186 using the A2 
tune~\cite{ATL-PHYS-PUB-2011-014} and the MSTW2008LO PDF set~\cite{Martin:2009iq}.
The events are reweighted to accurately reproduce the average number of interactions per bunch crossing in data.

All simulated samples are processed with a full ATLAS detector simulation~\cite{:2010wqa} based on
\textsc{Geant4}~\cite{Agostinelli:2002hh}.
The simulated events are reconstructed and analysed with the same analysis chain as for the data,
using the same trigger and event selection criteria discussed in Section \ref{sec:Selection}.

%% file: Objects.tex
Photons are reconstructed from clusters of energy deposits in the electromagnetic calorimeter measured in projective towers.
Clusters without matching tracks are classified as unconverted photon candidates.
A photon is considered as a converted photon candidate if it is matched to a pair of tracks that pass a  requirement on TRT-hits~\cite{photonreco} and form a vertex in the ID which is consistent with originating from a massless particle,
or if it is matched to a single track passing a TRT-hits requirement and has a first hit after the innermost layer of the pixel detector.
The photon energy is corrected by applying the energy scales measured with
$Z \rightarrow e^+ e^-$ decays~\cite{egamma2014}.
The trajectory of the photon is reconstructed using the longitudinal (shower depth) segmentation of the calorimeters and a constraint from the average collision point of the proton beams.
For converted photons, the position of the conversion vertex is also used if tracks
from the conversion have hits in the silicon detectors. 
Identification requirements are applied in order to reduce the contamination from $\pi^0$ or other neutral hadrons 
decaying to two photons. The photon identification is based on the profile of the energy deposits in the first  and second layers of the
 electromagnetic calorimeter. 
 Candidate photons are required to have $\pt > 10\GeV$, to satisfy the "loose" identification criteria defined in Ref.~\cite{PhId} 
  and to be within $\abseta < 2.37$. 
  Photons used in the event selection must additionally satisfy the "tight" identification criteria~\cite{PhId} and be isolated as 
follows. The energy in the calorimeters in a cone of size $\Delta R = \sqrt{(\Delta \eta)^2 + (\Delta \phi)^2} = 0.4$ around the 
cluster barycentre excluding the energy associated with the photon cluster is required to be less than $2.45\GeV + 0.022\pt^{\gamma}$, where $\pt^{\gamma}$ is the $\pt$ of the photon candidate. 
This cone energy is corrected for the leakage of the photon energy from the central core and for the
effects of pile-up~\cite{egamma2014}.

Electrons are reconstructed from clusters in the electromagnetic calorimeter matched to a track in the ID.
The criteria for their identification, and the calibration steps, are similar to those used for photons.
Electron candidates must satisfy the "medium" identification requirement of Ref.~\cite{egamma2014}.
Muons are identified either as a combined track in the MS and ID systems,
or as an ID track that, once extrapolated to the MS, is associated with at least one track segment in the MS.
Muon candidates must satisfy the "medium" identification requirement~\cite{Muons13TeV}.
The significance of the transverse impact parameter, defined as the transverse impact parameter $d_0$ divided by its estimated
uncertainty, $\sigma_{d_0}$, of tracks with respect to the primary vertex 
\footnote{The primary vertex is defined as  the vertex with the highest sum of the squared transverse momenta of its associated tracks. 
It is reconstructed from at least two associated tracks with $\pt > 0.4\GeV$.}
 is required to satisfy $|d_0|/\sigma_{d_0} <$ 5.0 for 
electrons and $|d_0|/\sigma_{d_0} <$ 3.0 for muons. The longitudinal impact parameter $z_0$ must be $|z_0|\sin\theta <$ 0.5 mm
for both electrons and muons.
Electrons are required to have $\pt > 7\GeV$ and $\abseta < 2.47$, while muons are 
required to have $\pt > 6\GeV$ and $\abseta  < 2.7$.
If any selected electron shares its inner detector 
track with a selected muon, the electron is removed and the muon is kept, in order to 
remove electron candidates coming from muon bremsstrahlung followed by photon conversion.

Jets are reconstructed using the anti-$k_{t}$ algorithm~\cite{Cacciari:2008gp, Cacciari:2005hq} with a radius parameter $R = $ 0.4 from clusters of energy deposits at the electromagnetic scale in the calorimeters.
A correction used to calibrate  the jet energy  to the scale of its constituent particles
\cite{Aad:2011he,JES2} is then applied.
In addition, jets are corrected for contributions from pile-up interactions~\cite{Aad:2011he}.
Candidate jets are required to have $\pt > 20\GeV$.
To suppress pile-up jets, which are mainly at low \pt, a jet vertex tagger~\cite{JVT}, based on tracking and vertexing information, 
is applied in jets with \pt\ $< 50\GeV$~  and $\abseta<2.4$. 
Jets used in the event selection are required to have \pt\ $>30\GeV$ and $|\eta| < 4.5$.
Hadronically decaying $\tau$ leptons are considered as jets as in the Run 1 analysis~\cite{PRD2015}.

To resolve ambiguities which can happen in object reconstruction,
an overlap removal procedure is performed in the following order.
If an electron lies within $\Delta R < 0.2$ of a candidate jet, the jet is removed
 from the event, while if an electron lies within $0.2 < \Delta R < 0.4$ of a  jet, the electron is removed.
Muons lying within $\Delta R < 0.4$ with respect to the remaining candidate jets
are removed, except if the number of tracks with $\pt > 0.5\GeV$  associated with the jet is less
than three. 
In the latter case, the jet is discarded and the muon kept. 
Finally if a candidate photon lies within $\Delta R < 0.4$ of a jet, the jet is removed.

The momentum imbalance in the transverse plane is obtained
from the negative vector sum of the reconstructed and calibrated physics objects, 
selected as described above, 
and is referred to as missing transverse momentum, \metv.
The symbol $\met$ is used to denote its magnitude.
Calorimeter energy deposits and tracks are associated with a reconstructed and identified high-$\pt$ object in a specific order: 
electrons with \pt$>7\GeV$, photons with \pt\ $>10\GeV$, and jets with \pt\ $>20\GeV$ \cite{Aad:2012re}.
Tracks from the primary vertex  not associated with any such objects ("soft term") are also taken into account in the \metv\ 
reconstruction \cite{MET_2012}. This track-based soft term is more robust against pile-up and provides a better \metv\ measurement in 
terms of resolution and scale than the calorimeter-based soft term used in Ref. \cite{PRD2015}.

Corrections are applied to the objects in the simulated samples to account for differences compared to data in object reconstruction, identification and isolation efficiencies for both the selected leptons and photons and for the vetoed leptons.

%% file: Event-selection.tex
The data were collected in $pp$ collisions at $\sqrt{s}=13 \TeV$  during 2015. 
The events for the analysis are recorded using a trigger requiring at least one photon candidate with an online $\pt$ threshold of 
$120\GeV$ passing "loose" identification requirements based on the shower shapes in the EM calorimeter as well as on the energy leaking into the hadronic calorimeter from the EM calorimeter~\cite{PhIsol}. 
Only data satisfying beam, detector and data quality criteria are considered.
The data used for the analysis correspond to an integrated luminosity of 3.2 \ifb. The uncertainty in the integrated luminosity is $\pm$5\%.
It is derived following a methodology similar to that detailed in Ref.~\cite{lumi2013}, from a preliminary calibration of the 
luminosity scale using $x$--$y$ beam-separation scans performed in August 2015. 

Quality requirements are applied to photon candidates in order to reject events containing photons arising from instrumental problems or
from non-collision background~\cite{photonreco}.
Beam-induced background is highly suppressed by applying the criteria described in Section~\ref{sec:BIB}.
In addition, quality requirements are applied to remove events containing candidate jets arising from detector noise and out-of-time energy 
deposits in the calorimeter from cosmic rays or other non-collision sources \cite{jetcleaning2}.
Events are required to have a reconstructed primary vertex. 

The criteria for selecting events in the signal region (SR) are optimized considering the discovery potential for  
 the simplified dark matter model.  This SR also provides good sensitivity to the other models described in Section~\ref{sec:intro}.
Events in the SR are required to have $\MET > 150\GeV$ and the leading photon has to satisfy the "tight" identification criteria, to 
have  $\pt^{\gamma}>150 \GeV$, $|\eta|<2.37$, excluding the calorimeter barrel/end-cap transition region $1.37<|\eta|<1.52$, and
to be isolated. 
With respect to the Run 1 analysis, a re-optimization was performed that leaded to the following changes: a higher threshold  for $\pt^{\gamma}$ ($150\GeV$ instead of $125\GeV$)
and a larger $\abseta$ region  ($\abseta <$  2.37 instead of 1.37) are used for the leading photon.
 It is required that the photon and \metv\ do not overlap in the azimuth: $\Delta\phi(\gamma,~$\metv$) > 0.4$.
Events with more than one jet or with a jet with $\Delta\phi(\jet,~$\metv$)<0.4$ are rejected.  
The remaining events with one jet are retained to increase the signal acceptance and reduce systematic uncertainties related to the modelling of initial-state radiation.
Events are required to have no electrons  or muons passing the requirements described in Section \ref{sec:Objects}.
The lepton veto mainly rejects $W/Z$ events with charged leptons in the final state.
For events satisfying these criteria, the efficiency of the trigger used in the analysis is $0.997^{+0.003}_{-0.008}$, as determined 
using a control sample of events selected with a \met\ trigger with a threshold of $70\GeV$.

The final data sample contains 264 events, of which 80 have a converted photon, and
170 and 94 events have zero and one jet, respectively.  

  The total number of events observed in the SR in data is compared with
   the estimated total number of events in the SR from SM backgrounds. 
   The latter is obtained from a simultaneous fit to various control regions
   (CR) defined in the following. Single-bin SR and CRs are considered 
   in the fit: no shape information within these regions is used.

%% file: Background.tex
The SM background to the $\gamma+\missET$ final state
 is dominated by the $Z(\rightarrow \nu \nu)\gamma$ process, where the photon is due to initial-state radiation.
Secondary contributions come from $W\gamma$  and $Z\gamma$ production with unidentified electrons, muons or with  hadronically decaying $\tau$ leptons.
There is also a contribution from  $W/Z$ production  where a lepton or an associated radiated jet is misidentified as a photon.
 In addition, there are smaller contributions from top-quark pair, diboson, $\gamma$+jets and multi-jet production.

All  background estimations are extrapolated from orthogonal data samples.
Control regions, built to be enriched in a specific background, are used to constrain the normalization of 
$W/Z\gamma$ and $\gamma$+jets backgrounds.
The normalization is obtained via a simultaneous likelihood fit~\cite{HistFitter} to the observed yields in all single-bin  CRs. 
 Poisson likelihood functions are used to model the expected event yields in all regions. The systematic uncertainties described in 
Section \ref{sec:syst} are treated as Gaussian-distributed nuisance parameters in the likelihood function.
 The fit in the CRs is performed to obtain the normalization factors for the $W\gamma$, $Z\gamma$ and $\gamma$+jets processes, 
which are then used to constrain background estimates in the SR. The same normalization factor is used for 
both $Z(\rightarrow \nu\nu)\gamma$ and $Z$ decaying to charged leptons in SR events.
  
The backgrounds due to fake photons from the misidentification of electrons or jets in $W/Z$+jets, top, diboson and multi-jet events are 
estimated using data-driven techniques based on studies of electrons and jets faking photons (see Sections \ref{sec:bElefake} and \ref{sec:bJetfake}).

\subsection{ $Z\gamma$ and $W\gamma$ backgrounds}
\label{sec:bVgamma}
  
   For the estimation of the $W/Z\gamma$ background, 
  three control regions are defined by selecting events with the same criteria used for the SR but inverting the lepton vetoes.
 In the first control region (1muCR) the $W\gamma$ contribution is enhanced by requiring the presence of a muon. The second and third control
 regions enhance the $Z\gamma$ background by requiring the presence of a pair of muons (2muCR) or electrons (2eleCR).
   In both 1muCR and 2muCR, to ensure that the $\missET$ spectrum is similar to the one in the SR, muons are treated as non-interacting particles in the $\missET$  reconstruction.
The same procedure is followed for electrons in the 2eleCR.
  In each case, the CR lepton selection follows the same requirements as the SR lepton veto, with the addition that the leptons must be isolated with "loose" criteria~\cite{Muons13TeV}.   
 In both the $Z\gamma$-enriched control regions,  the dilepton invariant mass $m_{\ell \ell}$ is required to be greater than $20 \GeV$. 
  The normalization of the dominant $Z\gamma$ background process is largely constrained by the event yields in the 2muCR and the 2eleCR.
The signal contamination in all CRs is negligible.  The expected fraction of signal events in the 1muCR is at the level of $0.15\%$. 
In the 2muCR and 2eleCR the contamination is zero due to the requirement of two leptons.

\subsection{$\gamma$+jets background}
\label{sec:bGammajet}

The $\gamma$+jets background in the signal region consists of events where the jet is poorly
reconstructed and partially lost, creating  fake $\MET$.
This background is suppressed by the large $\MET$ and the large jet--\metv\ azimuthal separation requirements.
It is estimated from simulated $\gamma$+jets events corrected with a 
normalization factor that is determined in a specific control region (PhJetCR), enriched in $\gamma$+jets events. 
This CR is defined with the same criteria as used for the SR, but requiring  $85\GeV <  \met < 110\GeV$ and 
azimuthal separation between the photon and \metv,  $\Delta\phi(\gamma,~$\metv$)$, to be smaller than 3, to minimize the contamination 
from signal events. 
The upper limit on the expected fraction of signal events in the
PhJetCR has been estimated to be at the level of 3\%.
 The extrapolation in \met\ of the gamma+jets background from the CR to the
  SR was checked in a validation region defined with higher \met\ 
 ($125 < \met < 250\GeV$) and requiring $\Delta\phi(\gamma,~$\metv$) < 3.0$; 
  no evidence of mismodeling was found.

\subsection{ Fake photons from misidentified electrons}
\label{sec:bElefake}
Contributions from processes in which an electron is misidentified as a photon are estimated by scaling yields from a sample of 
$e +\missET$ events by an electron-to-photon misidentification factor. This factor is measured with mutually exclusive samples of 
$e^+e^-$ and $\gamma+e$ events in data. To establish a pure sample of electrons, $m_{ee}$ and $m_{e\gamma}$ are both required to be 
consistent with the $Z$ boson mass to within $10\GeV$, and the $\missET$ is required to be smaller than $40\GeV$.
The  misidentification factor, calculated as the
ratio of the number of $\gamma+e$  to  the number of $e^+e^-$  events, is parameterized as a function of $\pt$  and pseudorapidity and it 
varies between $0.8\%$ and  $2.6\%$.
Systematic uncertainties from three different sources are added in quadrature: 
the difference between misidentification factors measured in data in two different windows around the $Z$ mass ($5\GeV$ and $10\GeV$),
the difference when measured in $Z(\rightarrow ee)$ MC events with the same method as used in 
data compared to using generator-level information, 
and the difference when measured in $Z(\rightarrow ee)$ and $W(\rightarrow e\nu)$ MC events using generator-level information.
Similar estimates are made for the three control regions with leptons, 
by applying the misidentification factor to events selected using the same 
criteria as used for these control regions but requiring an electron instead of a photon.
The estimated contribution of this background in the SR and the associated error are reported in Section~\ref{sec:result}. 

\subsection{ Fake photons from misidentified jets}
\label{sec:bJetfake}
Background contributions from events in which a jet is misidentified as a photon are estimated
using a sideband counting method~\cite{FakeJet_method}.
This method relies on counting photon candidates in four regions of a
two-dimensional space, defined by the transverse isolation energy and by
the quality of the identification criteria.
 A signal region (region A) is defined by photon candidates that are isolated with tight identification. 
Three background regions are defined, consisting of photon candidates which are either tight and 
non-isolated (region B), non-tight and isolated (region C) or non-tight and non-isolated (region D). 
The method relies on the fact that signal contamination in the three background regions
is small and that the isolation profile in the non-tight region is the 
same as that of the background in the tight region.
The number of background candidates in the signal region ($N_{\rm A}$) is calculated by taking the ratio
of the two non-tight regions ($N_{\rm C}/N_{\rm D}$) multiplied by the number 
of candidates in the tight, non-isolated region ($N_{\rm B}$).
This method is applied in all analysis regions: the SR and the four CRs.
The systematic uncertainty of the method is evaluated by varying the 
criteria of tightness and isolation used to define the four regions.
This estimate also accounts for the contribution from multi-jet events, which can mimic the $\gamma+\missET$ signature if one jet is misreconstructed as a photon and one or more of the other jets are poorly reconstructed, resulting in large $\MET$.
The estimated contribution of this background in the SR and the associated error are reported in Section~\ref{sec:result}.

\subsection{Beam-induced background}
\label{sec:BIB}

Muons from beam background can leave significant energy deposits in the calorimeters, mainly in the region at large $\abseta$,
and hence can lead to reconstructed fake photons.
These beam-induced fakes do not point back to the primary vertex, and the photon trajectory provides a powerful rejection criterion.
The $|z|$ position of the intersection of the extrapolated photon trajectory with the beam axis is required to be
smaller than 0.25 m, which rejects $98.5\%$ of these fake photons.
The residual beam background after the final event selection is found to 
be negligible, about $0.02\%$.

\subsection{ Final background estimation}

Background estimates in the SR are derived from a simultaneous fit to the four single-bin control regions 
(1muCR, 2muCR, 2eleCR and PhJetCR) in order to assess whether the observed SR 
yield is consistent with the background model. 
For each CR, the inputs to the fit are: 
the number of events seen in the data, 
the number of events expected from MC simulation for the $W/Z\gamma$ and $\gamma$+jets backgrounds, 
whose normalizations are free parameters, 
and the number of fake-photon events obtained from the data-driven techniques.
The fitted values of the normalization factors for $W\gamma$ and $Z\gamma$ are
 $k_{W\gamma}= 1.50 \pm 0.26$ and $k_{Z\gamma}= 1.19 \pm 0.21$, while 
 the normalization factor for the $\gamma$+jets background is $k_{\gammajets} = 0.98 \pm 0.28$. 
 The uncertainties include those from the various sources described in Section \ref{sec:syst}.
 The factor $k_{W\gamma}$ is large owing to the data--MC normalization difference in the 1muCR,
 which can potentially be reduced  using higher-order corrections for the $V\gamma$ cross sections~\cite{Vgamma_corr},
 which are not available for the selection criteria used here.

\begin{figure}[!htbp]
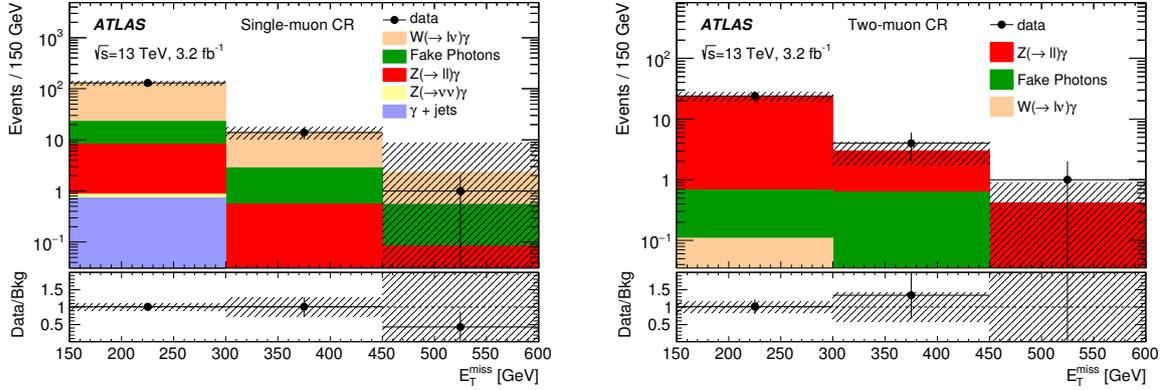
\centering
\includegraphics[width=0.49\linewidth]{{{fig_03a}}}
\includegraphics[width=0.49\linewidth]{{{fig_03b}}}
\caption{
Distribution of $\missET$, reconstructed treating muons as non-interacting particles, in the data and for the background in 
the 1muCR (left) and in the 2muCR (right). 
The total background expectation is normalized to the post-fit  result in each control region. 
Overflows are included in the final bin. 
The error bars are statistical, and the dashed band includes statistical and 
systematic uncertainties determined by a bin-by-bin fit. 
The lower panel shows the ratio of data to expected background event yields.}
\label{fig:auxiliary_1mu_2mu}
\end{figure}

\begin{figure}[!htbp]
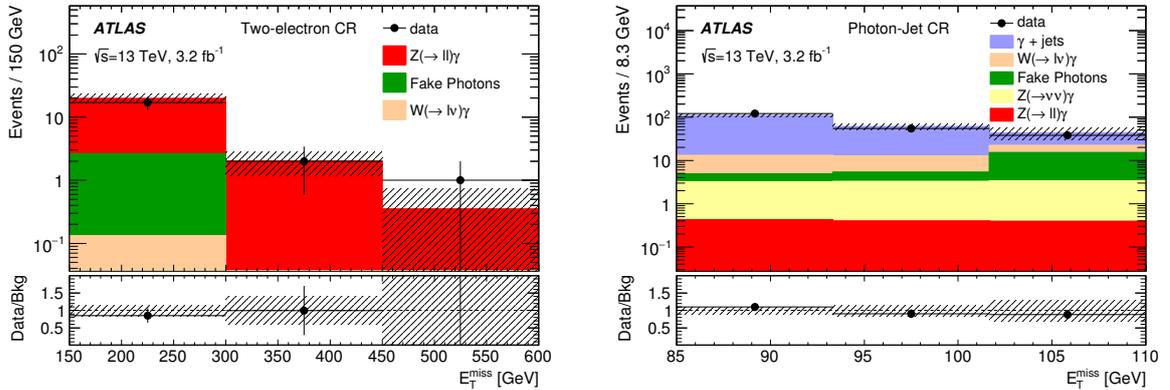
\centering
\includegraphics[width=0.49\linewidth]{{{fig_04a}}}
\includegraphics[width=0.49\linewidth]{{{fig_04b}}}
\caption{
Distribution of $\missET$ in the data and for the background in 
the 2eleCR, where $\missET$ is reconstructed treating electrons as non-interacting particles (left) 
and in the PhJetCR (right). 
The total background expectation is normalized to the post-fit  result in each control region. 
Overflows are included in the final bin for the left figure. 
The error bars are statistical, and the dashed band includes statistical and 
systematic uncertainties determined by a bin-by-bin fit. 
The lower panel shows the ratio of data to expected background event yields.}
\label{fig:auxiliary_2e_gammajets}
\end{figure}

Post-fit distributions of $\missET$ in the three lepton CRs  
and in the PhJetCR are shown in Fig.~\ref{fig:auxiliary_1mu_2mu} and Fig.~\ref{fig:auxiliary_2e_gammajets}.
These distributions illustrate the kinematics of the selected events. Their shape is not used in the simultaneous fit, which is performed on the single-bin CRs.

%% file: Results.tex
\begin{figure}[!htbp]\centering
\includegraphics[width=0.49\linewidth]{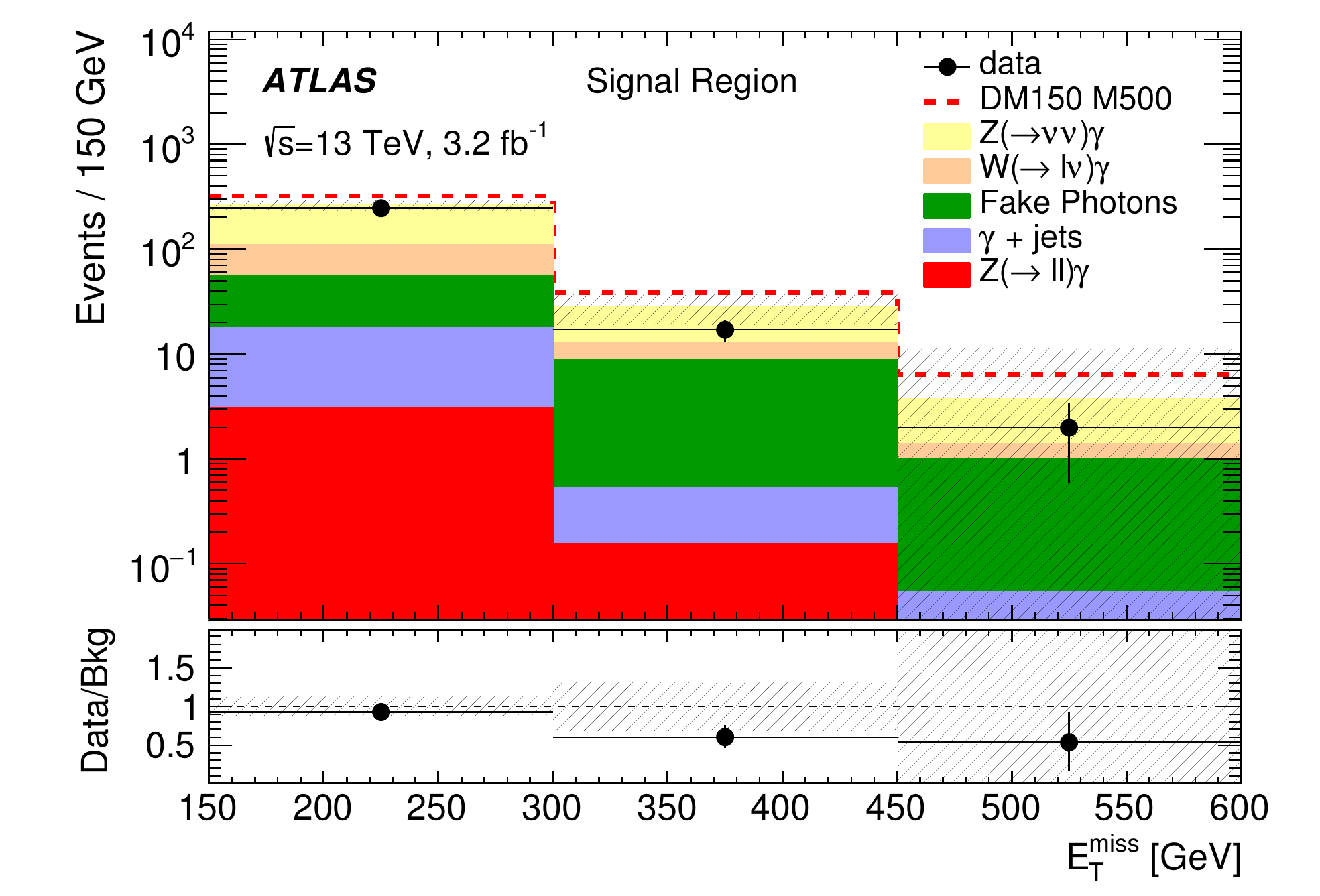}
\includegraphics[width=0.49\linewidth]{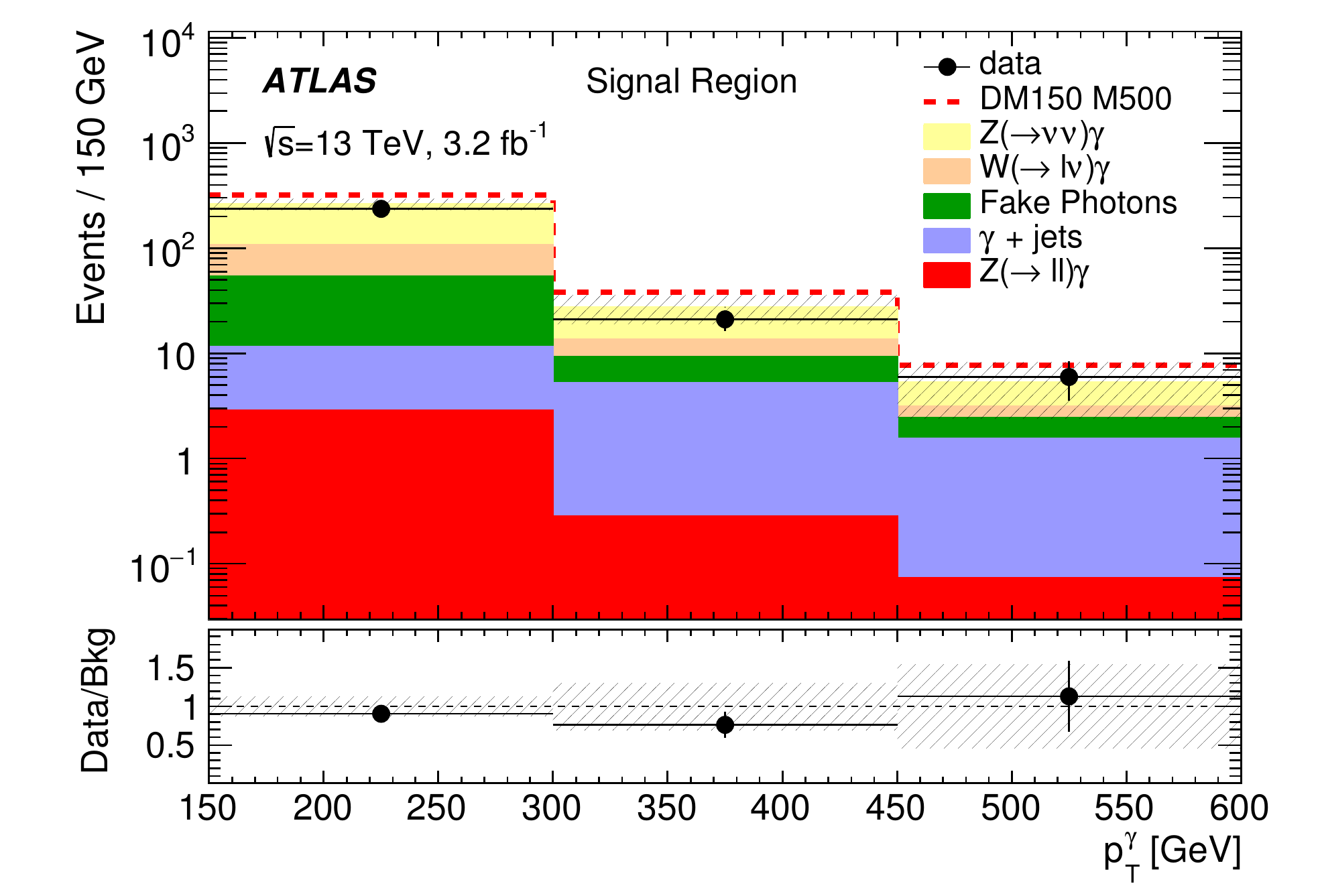}
\caption{
Distribution of $\missET$ (left) and photon $\pt$ (right)  in the signal region for data and for the background predicted from the fit in the CRs. 
Overflows are included in the final bin. 
The error bars are statistical, and the dashed band includes statistical and 
systematic uncertainties determined by a bin-by-bin fit. 
The expected yield of events from the simplified model with $m_{\chi} = 150\GeV$ and $m_{\rm med} = 500\GeV$ 
 is stacked on top  of the background prediction.
The lower panel shows the ratio of data to expected background event yields.
}
\label{fig:SR_event}
\end{figure}

Table \ref{tab:SR_yields} presents the observed number of events and the SM background 
predictions in the SR, obtained from the simultaneous fit to the single-bin CRs. 
The same numbers are also shown in the three lepton CRs and in the PhJetCR. 
The contribution from $W/Z\gamma$ with $W/Z$ decaying to $\tau$  includes both the leptonic and the hadronic $\tau$ decays, considered in this search as jets.  The fraction of $W(\rightarrow \tau\nu)$ and $Z(\rightarrow\tau\tau)$  with respect to  the total background corresponds to about  $12\%$ and $0.8\%$, respectively.  
The post-fit $\missET$ distribution and the photon $\pt$ distribution in the SR are shown in Fig.~\ref{fig:SR_event}.

\begin{table}[!htbp]
\begin {center}
\caption{Observed event yields in 3.2 \ifb~compared to  expected yields from SM
backgrounds in the signal region (SR) and in the four control regions (CRs), 
as predicted from the simultaneous  fit to all single-bin CRs.  The MC yields before
the fit are also shown.
The uncertainty includes both the statistical and systematic uncertainties described in
Section~\ref{sec:syst}.
The individual uncertainties can be correlated and do not necessarily 
add in quadrature to equal the total background uncertainty.}

\label{tab:SR_yields}
\begin{tabular}{c D{,}{\pm}{-1} D{,}{\pm}{-1} D{,}{\pm}{4.4} D{,}{\pm}{4.4}
D{,}{\pm}{4.4} }
\toprule
\toprule
& \multicolumn{1}{c}{ SR}& \multicolumn{1}{c}{ 1muCR}& \multicolumn{1}{c}{ 2muCR}&
\multicolumn{1}{c}{ 2eleCR}& \multicolumn{1}{c}{ PhJetCR}\\
\midrule
\multicolumn{1}{c}{ Observed events} & \multicolumn{1}{c}{264} & \multicolumn{1}{c}{
145} &  \multicolumn{1}{c}{ 29} &  \multicolumn{1}{c}{ 20} &  \multicolumn{1}{c}{
214} \\
\midrule

\multicolumn{1}{c}{Fitted Background}  & 295,34 & 145,12 &  27,4&23,3 &214,15\\
\midrule
\multicolumn{1}{c}{$Z(\rightarrow \nu\nu)\gamma$ }    & 171,29& 0.15,0.03 &0.00,0.00 &0.00,0.00
&8.6,1.4\\
\multicolumn{1}{c}{ $W(\rightarrow \ell\nu)\gamma$}           & 58, 9        & 119,17 &
0.14,0.04&0.11,0.03 &22,4\\
\multicolumn{1}{c}{$Z(\rightarrow \ell\ell)\gamma$}                     & 3.3,0.6        & 7.9,1.3 &
26,4&20,3 &1.2,0.2\\                      
\multicolumn{1}{c}{$\gammajets$ }             & 15, 4            & 0.7,0.5 &
0.00,0.00&0.03,0.03 &166,17\\
\multicolumn{1}{c}{ Fake photons from electrons}   & 22,18        & 1.7,1.5 &
0.05,0.05&0.00,0.00 &5.8,5.1\\  
\multicolumn{1}{c}{Fake photons from jets }   & 26,12           & 16,11 &
1.1,0.8&2.5,1.3 &9.9,3.1\\   
\midrule
\multicolumn{1}{c}{Pre-fit background}  & 249, 29 & 105,14 & 23,2 &19,2 &209,50\\

\bottomrule
\bottomrule
\end{tabular}
\end {center}
\end{table}

%% file: Systematics.tex
Systematic uncertainties in the background predictions in the SR are presented as percentages of the total background prediction.
This prediction is obtained from the simultaneous fit to all single-bin CRs, which provides constraints on many sources of systematic uncertainty,
as the normalizations of the dominant background processes are fitted parameters. 
The dominant systematic uncertainties are summarised in Table~\ref{tab:syst}. 

The total background prediction uncertainty, including systematic and statistical contributions,  is approximately 11\%, dominated by the  statistical uncertainty in the control regions, which amounts to approximately 9\%.
The largest relative systematic uncertainty of 5.8\% is due to the electron fake rate. This is  mainly driven by the small number of 
events available for the  estimation of the electron-to-photon misidentification factor yielding a precision of  30--100\%, depending 
on \pt\ and $\eta$.
PDF uncertainties have an impact on  the $V\gamma$ samples in each region but the effect on normalization is largely absorbed in the 
fit. 
They are evaluated 
following the prescriptions of the PDF group recommendations~\cite{PDF4LHC} 
and using a reweighting procedure implemented in the LHAPDF Tool~\cite{LHAPDF}. 
These uncertainties contribute 2.8\% to the background prediction uncertainty affecting mainly 
the $Z(\rightarrow \nu\nu) \gamma$ background.
The uncertainty on the jet fake rate contributes a relative uncertainty of 2.4\% and affects mainly 
 the normalization of  \wgamma\  background, while 
the uncertainty on the muon reconstruction and isolation efficiency gives a relative uncertainty of  1.5\% and mainly affects the \zgamma\ background.
 Finally the uncertainty on the jet energy resolution accounts for 1.2\% of the uncertainty
 and the most affected background is \gammajets.  
After the fit,  the uncertainty on the luminosity \cite{lumi2013} is found to have a negligible impact on the background estimation.

For the signal-related systematics, the PDF uncertainties are evaluated in the 
same way described above for the background samples, while QCD scale 
uncertainties are evaluated by varying the renormalization and 
factorization  scales by factors 2.0 and 0.5 with respect to the 
nominal values used in the MC generation. The uncertainties due to the 
choice of underlying-event tune used with $\pythiaeight$ are 
computed by generating MC samples with the alternative underlying-event 
tunes described in Ref.~\cite{ATL-PHYS-PUB-2014-021}.

\begin{table}[!htbp]
\centering
\caption{
Breakdown of the dominant systematic uncertainties in the background estimates. 
The uncertainties are given relative to the expected total background yield. 
The individual uncertainties can be correlated and do not necessarily 
add in quadrature to equal the total background uncertainty.
 }
 
\label{tab:syst}
\begin{tabular}{lr}
\hline\hline
Total background & 295 \\ 
Total background uncertainty  & $11\%$\\ 
\hline
Electron fake rate                      & $5.8\%$     \\
PDF uncertainties  & $2.8\%$ \\
Jet fake rate                      & $2.4\%$    \\                      
Muons reconstruction/isolation efficiency    & $1.5\%$       \\  
Electrons reconstruction/identification/isolation efficiency &  $1.3\%$         \\
Jet energy resolution~\cite{JER}  & $1.2\%$ \\
Photon energy scale     & $0.6\% $          \\
\metv\ soft term scale and resolution & $0.4\%$ \\   
Photon energy resolution     & $0.2\% $          \\        
Jet energy scale~\cite{JES2}  & $0.1\%$ \\
\hline
\hline
\end{tabular}
\end{table}

%% file: Interpretation.tex
The 264 events observed in data are consistent with the prediction of $ 295\pm 34$ events from SM backgrounds.
The results are therefore interpreted in terms of exclusion limits in models that would produce an excess of  $\gamma+\missET$ events.
 Upper bounds are calculated using a one-sided profile likelihood ratio and the $CL_S$ 
technique~\cite{Read:2002hq,Junk:1999kv}, evaluated using the asymptotic approximation~\cite{Cowan:2010js}.
The likelihood fit includes both the SR and the CRs.

Limits on the fiducial cross section of a potential signal beyond the SM, defined
as the product of the cross section  times the fiducial acceptance $A$, are
provided. These limits can be extrapolated within some approximations to models
producing $\gamma+\missET$  events once $A$ is known. 
The value of $A$ for a particular model is 
computed by applying the same selection criteria as in the SR but at the 
particle level; in this computation \metv\ is given by the vector sum of 
the transverse momenta of all invisible particles. 
 The value of $A$ is 0.43--0.56 (0.4)  for the DM (ADD) samples generated for this search 
 following the specifications given in Section \ref{sec:simsample}.
 The limit is computed by dividing the limit on the visible 
cross section $\sigma\times A \times \epsilon$ by the fiducial 
reconstruction efficiency $\epsilon$. The latter is conservatively taken 
to be 78\%, corresponding to the lowest efficiency found in the ADD and 
DM models studied here, for which the efficiency ranges from 78\% to 
91\%. 
 The observed (expected) upper limits on the fiducial cross section 
$\sigma \times A$ for the production of $\gamma+\missET$ events are 
17.8  (25.5) fb at 95\% confidence level (CL) and 14.6 (21.7) fb at 90\% CL.
The observed upper limit at 95\% CL would be 15.3 fb using the largest efficiency value
of 91\%.

When placing limits on specific models, the signal-related systematic 
uncertainties calculated as described in Section \ref{sec:syst}
affecting $A \times \epsilon$ (PDF, scales, initial- and 
final-state radiation) are included in the statistical analysis,  while 
the uncertainties affecting the cross section (PDF, scales) are 
indicated as bands around the observed limits and written as 
$\sigma_{\textrm{theo}}$. 

\begin{figure}
    \centering
	\includegraphics[width=0.54\textwidth]{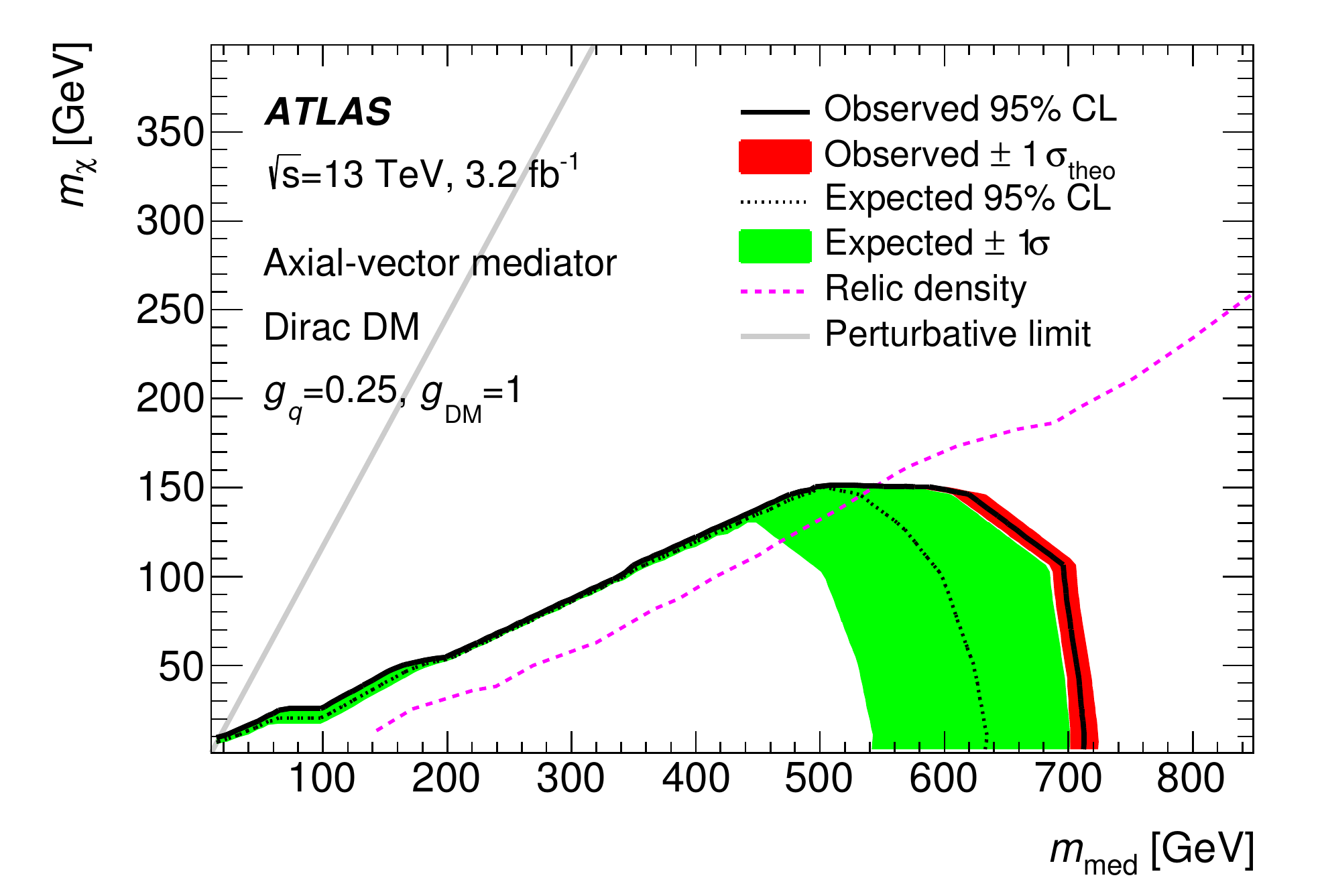}
	\caption{The observed and expected 95\% CL exclusion limit for a simplified model of dark matter production 
	involving an axial-vector operator, Dirac DM and couplings $g_q$ = 0.25 and $g_{\chi}$ = 1  
	 as a function of  the dark matter mass $m_{\chi}$ and the  axial-mediator mass $m_{\rm  med}$. 
	 The plane under the limit curves is excluded.
	 The region on the left is excluded by the perturbative limit.
	  The relic density curve \cite{WMAP_new} is also shown.}
	\label{fig:dmA_exclusion}
\end{figure}

Simplified models with explicit mediators are
robust for all values of the momentum transfer 
$Q_{\textrm{tr}}$~\cite{dmforum}. 
For the simplified model with an axial-vector mediator, Fig.~\ref{fig:dmA_exclusion} shows the observed and expected contours 
corresponding to a 95\% CL exclusion as a function of $m_{\rm med}$ and  $m_{\chi}$ for $g_q$ = 0.25 and $g_{\chi}$ =1. The region of the plane under the limit curves is excluded.
 The region not allowed due to perturbative unitarity violation is to the left of the line defined by 
 $m_{\chi} = \sqrt{\pi/2}m_{\rm med}$~\cite{Perturb}.
 The line corresponding to the DM thermal relic abundance \cite{WMAP_new} is also indicated in the figure. 
The search excludes mediator masses below $710\GeV$ for $\chi$ masses below $150\GeV$. 
 
\begin{figure}
    \centering
	\includegraphics[width=0.54\textwidth]{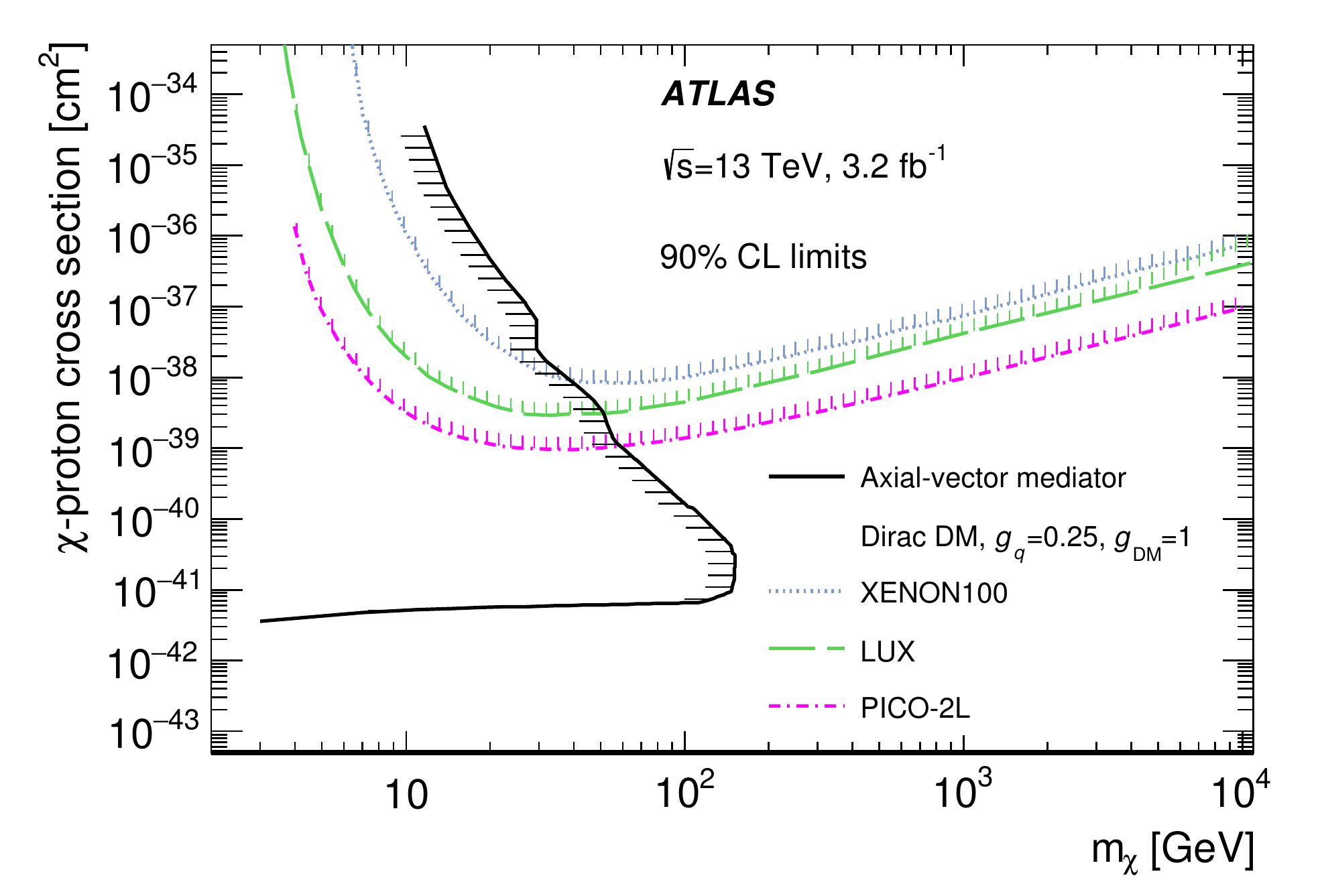}
	\caption{The 90\% CL exclusion limit on the $\chi$--proton scattering cross section in a simplifed model of dark matter production 
	involving an axial-vector operator, Dirac DM and couplings $g_q$ = 0.25 and $g_{\chi}$ = 1
        as a function of  the dark matter mass $m_{\chi}$.
         Also shown are results from three direct dark matter search experiments~\cite{PICO, XENON100_new, LUX_new}.}
	\label{fig:cont_dmA}
\end{figure}

 Figure~\ref{fig:cont_dmA} shows the contour corresponding to a 90\% CL exclusion translated to the $\chi$--proton scattering cross section vs. $m_\chi$ plane. 
 Bounds on the $\chi$-proton cross section are obtained following the 
procedure described in Ref. \cite{dmrecomm}, assuming that the axial-vector mediator 
with couplings $g_q = 0.25$ and $g_{\chi} = 1.0$ is solely responsible for both 
collider $\chi$ pair production and for $\chi$-nucleon scattering.
  In this plane a comparison with the result from direct DM searches~\cite{PICO, XENON100_new, LUX_new} is possible. 
The search provides stringent limits on the scattering cross section at the order of $10^{-41} {\rm cm^2}$ up to $m_{\chi}$ masses of about 
$150\GeV$. The limit placed in this search extends to arbitrarily low values of $m_{\chi}$, as the acceptance at lower mass values 
is the same as the one at the lowest $m_{\chi}$ value shown here.

\begin{figure}\centering
 	\includegraphics[width=0.54\textwidth]{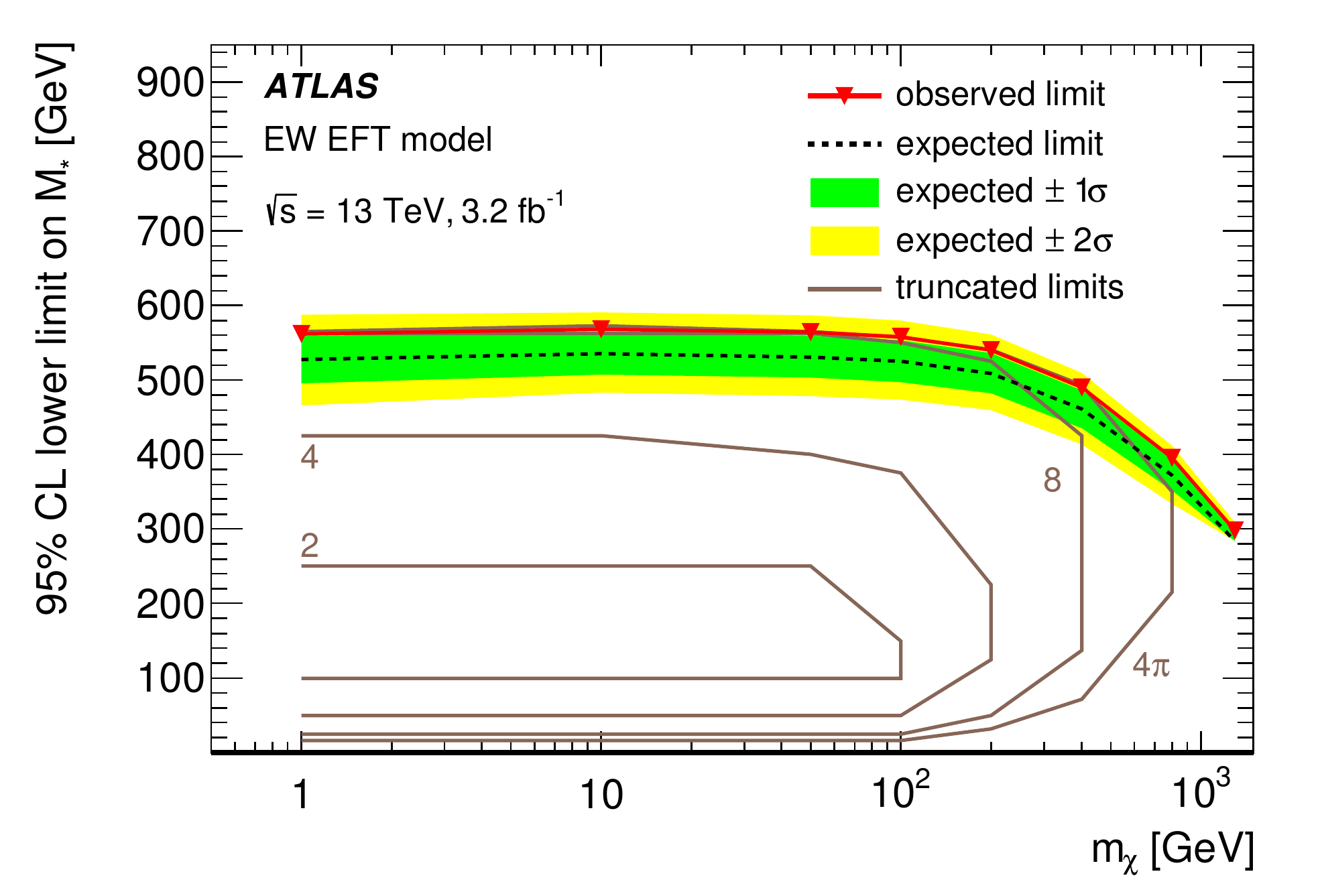}
	\caption{
	The observed and expected 95\% CL limits on $M_{*}$ for a  dimension-7 operator EFT model with a contact interaction of type $\gamma\gamma\chi\chi$  as a function of dark matter mass $m_{\chi}$.
Results  where EFT truncation is applied are also shown, assuming representative coupling values of 2, 4, 8 and 4$\pi$.
	 }
   \label{fig:Plot_schannel_Limit}
\end{figure}

In the case of the model of $\gamma\gamma\chi\bar{\chi}$ interactions, lower limits are placed on the effective 
mass scale $\Mstar$ as a function of $m_{\chi}$, as shown in Fig.~\ref{fig:Plot_schannel_Limit}. 
The EFT is not always valid, so a truncation procedure is applied~\cite{Truncation}. 
In this procedure, the scale at which the EFT description becomes invalid ($M_{\text{cut}}$) 
is assumed to be related to $M_{*}$ through  $M_{\text{cut}}=g^{*}M_{*}$, 
where $g^{*}$ is the EFT coupling. Events having a centre-of-mass energy larger than
 $M_{\text{cut}}$ are removed and the limit is recomputed. 
The effect of the truncation for various representative values of $g^{*}$ 
is shown in Fig.~\ref{fig:Plot_schannel_Limit}: for the maximal coupling value of 4$\pi$, the truncation
 has almost no effect; for lower coupling values, the exclusion limits are confined 
 to a smaller area of the parameter space, and no limit can be set for a coupling value of unity. 
For very low values of $M_{*}$, most events would fail the centre-of-mass 
energy truncation requirement, therefore, the truncated limits are not 
able to exclude very low $M_{*}$ values.
The search excludes model values of $M_*$ up to $570\GeV$ and  
effects of truncation for various coupling values are shown in the figure.

\begin{figure}[!htbp]\centering
 	\includegraphics[width=0.54\textwidth]{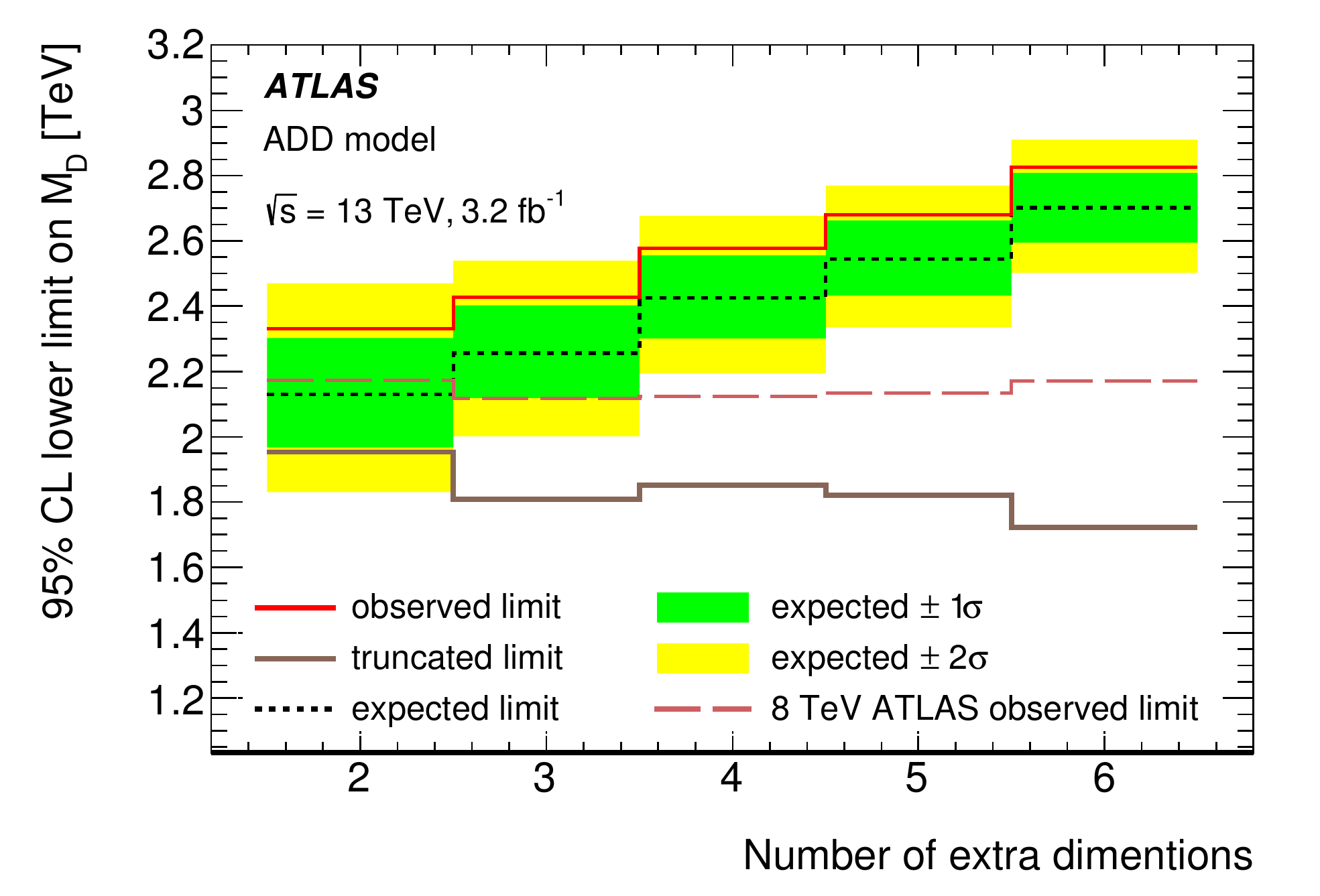}
	\caption{The observed and expected 95\% CL lower limits on the mass scale $M_{\rm D}$ in the ADD models of large extra 
                 dimensions, for several values of the number of extra dimensions.  
                 The untruncated limits from the search of $8\TeV$ ATLAS data~\cite{PRD2015} are shown for comparison. 
                 The limit with truncation is also shown.
                }
   \label{fig:ADD_Limit}
\end{figure}

In the ADD model of LED,  the observed and expected 95\% CL lower limits on  the fundamental Planck mass $M_{\rm D}$  for various values of $n$ are shown in Fig.~\ref{fig:ADD_Limit}. 
The values of $M_{\rm D}$ excluded at 95\% CL are larger for larger $n$ values: this is explained by the increase of the cross section at the centre-of-mass energy of $13\TeV$ with increasing $n$, which is an expected 
behaviour for values of $M_{\rm D}$ which are not large with respect to the centre-of-mass energy.
Results incorporating truncation in the phase-space region where the model implementation is not valid are also shown. This consists in suppressing the graviton production cross section by a factor $M_{\rm D}^{4}/s^2$ in events with centre-of-mass energy 
$\sqrt{s} > M_{\rm D}$. 
The procedure is repeated iteratively with the new truncated limit until it 
converges, i.e., until the difference between the new truncated limit 
and the one obtained in the previous iteration differ by less than 
0.1$\sigma$.
It results in a decrease of the 95\% CL limit on $M_D$.
The search sets limits that are more stringent than those from LHC Run 1,  
excluding $M_{\rm D}$ up to about $2.3 \TeV$ for $n$ = 2 and up to $2.8\TeV$  for $n$ = 6;
 the limit values are reduced by 20 to 40\% depending on $n$ when applying a truncation procedure.

%% file: Conclusion.tex
Results are reported on a search for new phenomena in events with a high-$\pt$ photon and large missing transverse momentum in $pp$ 
collisions at $\sqrt{s}=13\TeV$ at the LHC, using data collected by the ATLAS experiment corresponding to an integrated luminosity of 3.2 \ifb. 
 The observed data are consistent with the Standard Model expectations.
  The observed (expected) upper limits on the fiducial cross section for the production of events with a photon and large 
  missing transverse momentum  are 
  17.8  (25.5) fb at 95\% CL and 14.6 (21.7) fb at 90\% CL.
 For the simplified DM model considered, the search excludes 
mediator masses below $710\GeV$ for  $\chi$ masses below $150\GeV$. 
For the EW-EFT model values of $M_*$ up to $570\GeV$ are excluded and the 
effect of truncation for various coupling values is reported.
For the ADD model the search sets limits that are more stringent than in the Run 1 data search, 
excluding $M_{\rm D}$ up to about $2.3 \TeV$ for $n$ = 2 and up to $2.8\TeV$  for $n$ = 6;
 the limit values are reduced by 20--40\% depending on $n$ when applying a truncation procedure.

%% file: Acknowledgements.tex

We thank CERN for the very successful operation of the LHC, as well as the
support staff from our institutions without whom ATLAS could not be
operated efficiently.

We acknowledge the support of ANPCyT, Argentina; YerPhI, Armenia; ARC, Australia; BMWFW and FWF, Austria; ANAS, Azerbaijan; SSTC, Belarus; CNPq and FAPESP, Brazil; NSERC, NRC and CFI, Canada; CERN; CONICYT, Chile; CAS, MOST and NSFC, China; COLCIENCIAS, Colombia; MSMT CR, MPO CR and VSC CR, Czech Republic; DNRF and DNSRC, Denmark; IN2P3-CNRS, CEA-DSM/IRFU, France; GNSF, Georgia; BMBF, HGF, and MPG, Germany; GSRT, Greece; RGC, Hong Kong SAR, China; ISF, I-CORE and Benoziyo Center, Israel; INFN, Italy; MEXT and JSPS, Japan; CNRST, Morocco; FOM and NWO, Netherlands; RCN, Norway; MNiSW and NCN, Poland; FCT, Portugal; MNE/IFA, Romania; MES of Russia and NRC KI, Russian Federation; JINR; MESTD, Serbia; MSSR, Slovakia; ARRS and MIZ\v{S}, Slovenia; DST/NRF, South Africa; MINECO, Spain; SRC and Wallenberg Foundation, Sweden; SERI, SNSF and Cantons of Bern and Geneva, Switzerland; MOST, Taiwan; TAEK, Turkey; STFC, United Kingdom; DOE and NSF, United States of America. In addition, individual groups and members have received support from BCKDF, the Canada Council, CANARIE, CRC, Compute Canada, FQRNT, and the Ontario Innovation Trust, Canada; EPLANET, ERC, FP7, Horizon 2020 and Marie Sk{\l}odowska-Curie Actions, European Union; Investissements d'Avenir Labex and Idex, ANR, R{\'e}gion Auvergne and Fondation Partager le Savoir, France; DFG and AvH Foundation, Germany; Herakleitos, Thales and Aristeia programmes co-financed by EU-ESF and the Greek NSRF; BSF, GIF and Minerva, Israel; BRF, Norway; Generalitat de Catalunya, Generalitat Valenciana, Spain; the Royal Society and Leverhulme Trust, United Kingdom.

The crucial computing support from all WLCG partners is acknowledged
gratefully, in particular from CERN and the ATLAS Tier-1 facilities at
TRIUMF (Canada), NDGF (Denmark, Norway, Sweden), CC-IN2P3 (France),
KIT/GridKA (Germany), INFN-CNAF (Italy), NL-T1 (Netherlands), PIC (Spain),
ASGC (Taiwan), RAL (UK) and BNL (USA) and in the Tier-2 facilities
worldwide.

%% file: atlas_authlist.tex
\begin{flushleft}
{\Large The ATLAS Collaboration}

\bigskip

M.~Aaboud$^\textrm{\scriptsize 136d}$,
G.~Aad$^\textrm{\scriptsize 87}$,
B.~Abbott$^\textrm{\scriptsize 114}$,
J.~Abdallah$^\textrm{\scriptsize 65}$,
O.~Abdinov$^\textrm{\scriptsize 12}$,
B.~Abeloos$^\textrm{\scriptsize 118}$,
R.~Aben$^\textrm{\scriptsize 108}$,
O.S.~AbouZeid$^\textrm{\scriptsize 138}$,
N.L.~Abraham$^\textrm{\scriptsize 150}$,
H.~Abramowicz$^\textrm{\scriptsize 154}$
% ...
\bigskip
\\
% ...
$^{12}$ Institute of Physics, Azerbaijan Academy of Sciences, Baku, Azerbaijan\\
% ...
$^{65}$ University of Iowa, Iowa City IA, United States of America\\
% ...
$^{87}$ CPPM, Aix-Marseille Universit{\'e} and CNRS/IN2P3, Marseille, France\\
% ...
$^{108}$ Nikhef National Institute for Subatomic Physics and University of Amsterdam, Amsterdam, Netherlands\\
% ...
$^{114}$ Homer L. Dodge Department of Physics and Astronomy, University of Oklahoma, Norman OK, United States of America\\
% ...
$^{118}$ LAL, Univ. Paris-Sud, CNRS/IN2P3, Universit{\'e} Paris-Saclay, Orsay, France\\
% ...
$^{136}$ $^{(a)}$ Facult{\'e} des Sciences Ain Chock, R{\'e}seau Universitaire de Physique des Hautes Energies - Universit{\'e} Hassan II, Casablanca; $^{(b)}$ Centre National de l'Energie des Sciences Techniques Nucleaires, Rabat; $^{(c)}$ Facult{\'e} des Sciences Semlalia, Universit{\'e} Cadi Ayyad, LPHEA-Marrakech; $^{(d)}$ Facult{\'e} des Sciences, Universit{\'e} Mohamed Premier and LPTPM, Oujda; $^{(e)}$ Facult{\'e} des sciences, Universit{\'e} Mohammed V, Rabat, Morocco\\
% ...
$^{138}$ Santa Cruz Institute for Particle Physics, University of California Santa Cruz, Santa Cruz CA, United States of America\\
% ...
$^{150}$ Department of Physics and Astronomy, University of Sussex, Brighton, United Kingdom\\
% ...
$^{154}$ Raymond and Beverly Sackler School of Physics and Astronomy, Tel Aviv University, Tel Aviv, Israel\\
% ...
$^{a}$ Also at Department of Physics, King's College London, London, United Kingdom\\
$^{b}$ Also at Institute of Physics, Azerbaijan Academy of Sciences, Baku, Azerbaijan\\
$^{c}$ Also at Novosibirsk State University, Novosibirsk, Russia\\
$^{d}$ Also at TRIUMF, Vancouver BC, Canada\\
$^{e}$ Also at Department of Physics {\&} Astronomy, University of Louisville, Louisville, KY, United States of America\\
% ...
\end{flushleft}
